\documentstyle[12pt]{article}
\begin{document}
\newcommand{\cinf}{${\cal C}^\infty(M^0_q)\;$}
\newcommand{\re}{\mbox{Re}\,}
\newcommand{\im}{\mbox{Im}\,}
\newcommand{\beq}{\begin{equation}}
\newcommand{\eeq}[1]{\label{#1}\end{equation}}
\newcommand{\bea}{\begin{eqnarray}}
\newcommand{\eea}[1]{\label{#1}\end{eqnarray}}
\newcommand{\dirac}{/\!\!\!\partial}
\newcommand{\Dirac}{/\!\!\!\!D}
%% Less or approx and greater or approx:
\def\lequiv{\raise 0.4ex \hbox{$<$} \kern -0.8 em \lower 0.62 ex \hbox{$\sim$}
}
\def\gequiv{\raise 0.4ex \hbox{$>$} \kern -0.7 em \lower 0.62 ex \hbox{$\sim$}
}
\begin{titlepage}
\begin{center}
\hfill hep-th/9505187 \\
\hfill NYU-TH.95/05/03
\vskip .4in
{\large\bf On the Existence of States Saturating the Bogomol'nyi Bound
in N=4 Supersymmetry}
\end{center}
\vskip .4in
\begin{center}
{Massimo Porrati\footnotemark}
\vskip .4in
{\it Department of Physics, New York University\\
4 Washington Pl., New York NY 10003, USA.}
\footnotetext{On leave of absence from I.N.F.N., sez. di Pisa, Pisa, Italy.
E-mail: porrati@mafalda.nyu.edu}
\end{center}
\vskip .4in
\begin{center} {\bf ABSTRACT} \end{center}
\begin{quotation}
\noindent
We give an argument showing that in N=4 supersymmetric gauge theories
there exists at least one bound state saturating the
Bogomol'nyi bound with electric charge $p$ and magnetic charge $q$, for
each $p$ and $q$ relatively prime, and we comment on the uniqueness of such
state.
This result is a necessary condition for the existence of an exact S-duality
in N=4 supersymmetric theories.
 \end{quotation}
\vfill
\end{titlepage}
\noindent
A mounting body of evidence~\cite{misc,sen1,sen2}
supports the conjecture that N=4
rigid supersymmetric theories possess an $SL(2,Z)$ duality. This $SL(2,Z)$
contains as a subgroup the strong-weak duality of Montonen and
Olive~\cite{mo,osb,gno}.

Let us confine our attention to an N=4 theory with
gauge group $SU(2)$ spontaneously broken to $U(1)$, for simplicity.
In this case the duality acts by fractional transformations on the complex
number $S=\theta/2\pi + i4\pi/g^2$, which combines together the theta angle
$\theta$ and the gauge coupling constant $g$
\beq
S \rightarrow {aS + b\over cS +d},\;\;\; a,b,c,d \in Z,\;\;\; ad-bc=1.
\eeq{1}
The meaning of duality is that theories whose coupling constants are
related by a transformation of type~(\ref{1}) are physically equivalent. In
particular, the stable physical states of the two theories are the same,
up to an eventual relabeling of them. Among the physical states of the
theory there is a special subset: those that saturate the Bogomol'nyi
bound~\cite{bog}. These states belong to short multiplets of N=4 and thus
their mass is non renormalized~\cite{wo}. By denoting with $p,q\in Z$ the
electric and magnetic charge of one such state,
its mass is given by the following formula~\cite{wo,sen1}
\beq
M^2(p,q)= {4\pi V^2\over \im S}( p^2 +2\re S pq + |S|^2q^2).
\eeq{2}
Here $V$ is a constant independent of $p$, $q$ and $S$.

Not all assignments of $p$ and $q$ need give a stable state. Indeed, a $(p,q)$
state may decay in a pair $[(p_1,q_1),(p_2,q_2)]$ $p_1+p_2=p$, $q_1+q_2=q$,
whenever
\beq
M(p,q)\geq M(p_1,q_1) + M(p_2,q_2).
\eeq{3}
Since eq.~(\ref{2}) defines a positive-definite scalar product ($\im S>0$),
the triangular inequality holds: $M(p,q)\leq M(p_1,q_1) + M(p_2,q_2)$,
with the equality attained when $p_1/q_1=p_2/q_2=p/q$. The charges are
integers, thus this is possible only when $p$ and $q$ have a common divisor
\beq
p=Nm,\;\; q=Nn,\;\;\; N,n,m \in Z.
\eeq{4}

This argument, due to Sen~\cite{sen1,sen2}, does not guarantees that states
obeying the above equations actually decay, but it does shows that all
Bogomol'nyi states with $p,q$ relatively prime are stable.
Moreover, Bogomol'nyi states are transformed by S-duality. If this duality
exists, eq.~(\ref{2}) must be invariant under the $SL(2,Z)$
transformation~(\ref{1}), together with a relabeling of $p$ and $q$, which
turns out to be
\beq
p\rightarrow ap -bq, \;\;\; q \rightarrow -cp + dq.
\eeq{5}
If S-duality holds, eq.~(\ref{5}) maps the multiplet with $p=1,q=0$ (the
electrically charged vector multiplet of the $SU(2)$ theory, corresponding
to a broken generator of the gauge group) into states with $p=a$, $q=-c$.
These states must also be stable, by consistency. This property holds
because the constraints $ad-bc=1$, $a,b,c,d\in Z$ imply that $p$ and $q$ are
relatively prime.

Stable states with $p=0,1$ and $q=0,1$ are well known: they are,
respectively, the
elementary charged states of the $SU(2)$ gauge theory, the BPS monopole,
and the dyon. A crucial test of S-duality is to verify
whether {\em all} states with
$p,q$ relatively prime do indeed exist with the right multiplicity.
The case $q=2$, $p$ odd integer, has been studied in~\cite{sen2}. The
topological aspects of the generalization to arbitrary $q$ have been
demonstrated by Segal~\cite{se}.

Purpose of this paper is to show that for all $p,q$ relatively prime, there
exists at least one N=4 (short) multiplet of states whose mass saturates
the Bogomol'nyi bound~(\ref{2}), and to propose a possible strategy to prove
its uniqueness.

In~\cite{sen2}, a general strategy for finding Bogomol'nyi
states with arbitrary electric and magnetic charge has also been proposed.
Let us resume the features of that argument that we shall need in this
paper.

Let us put as in~\cite{sen2} $\theta=0$, the general case being (probably)
reconducible to this one once the Witten phenomenon~\cite{w} is taken into
account.
Let us also notice that it is sufficient to study our N=4 theory in the
small-coupling constant (semiclassical) regime: $g\rightarrow 0$. Indeed,
Bogomoln'yi states with $p,q$ relatively prime are stable for each value of
$g$ i.e. they cannot disappear (decay) as the coupling constant is
adiabatically switched off, or turned on.

In the semiclassical limit, the dynamics of states with magnetic charge $q$
is determined as follows~\cite{gm}. One looks for {\em static}
solutions of the classical equations of motion of the N=4 $SU(2)$ theory,
with magnetic charge $q>0$ (the case $q<0$ is analogous), and mass given by
eq.~(\ref{2}) $M= (4\pi/g)V q$. They depend on several bosonic and
fermionic moduli, that we denote collectively by $m$.
Let us call these solutions $\phi^o(\vec{x},m)$.
We may then decompose the most general N=4 field (better, multiplet of fields),
denoted by $\phi(t,\vec{x})$, as
\beq
\phi(x)\equiv \phi(t,\vec{x})=\phi^o(\vec{x},m(t)) + \delta^T\phi(t,\vec{x}).
\eeq{5'}
Note that the moduli have been given a time dependence. If one expands the
N=4 action, $S[\phi(t,\vec{x})]$,
up to quadratic terms (higher terms are irrelevant for $g\rightarrow 0$) one
finds that the linear terms are absent because $\phi^o$ is a solution of the
classical equations of motion, and the action reads
\beq
S[\phi(t,\vec{x})]= S[\phi^o(\vec{x},m(t))] +
\int d^4x d^4y\delta^T\phi(x) {\delta^2 S\over
\delta\phi(x)\delta\phi(y)}\delta^T\phi(y)
\eeq{5''}
The fluctuations $\delta^T\phi(x)$ are transverse, i.e. they obey
\beq
\int d^3 x {\delta^2 S\over \delta m(t) \delta \phi (x)}\delta^T\phi(x)=0,
\;\;\; \forall t
\eeq{5'''}
The quantization of the classical theory with lagrangian~(\ref{5''}) gives
a hamiltonian of the form
\beq
H= H^o[m, -i\partial_m] +
H^T[\delta^T\phi(\vec{x}),\delta^T\pi(\vec{x}),m].
\eeq{5''''}
The ``transverse hamiltonian,'' $H^T$, does not depend on $-i\partial_m$
because of eq.~(\ref{5'''}), thus $m$ enters in $H^T$ as an external
parameter.
It can be shown that upon quantization the energy levels of $H^T$ are
$(4\pi/g)Vq+{\cal O}(1)$, and that by supersymmetry the ground state energy of
$H^T$ is independent of $m$ and equal to $(4\pi/g)Vq$.
If a static
Bogomol'nyi state with electric charge $p$ exists, instead, its energy is
$(4\pi/g)Vq+ (g^3/8\pi q)Vp^2 + {\cal O}(g^5)$. Such state, therefore, is a
product $\psi(m)\otimes \Psi^0$, where $\Psi^0$ denotes
the ground state of $H^T$. $\Psi^0$
depends on $m$, but, as usual in the $g\rightarrow 0$ limit (a.k.a. the
Born-Oppenheimer approximation), this dependence is negligible; thus,
$\psi(m)$ is a {\em normalizable} eigenstate of the {\em quantum mechanical}
hamiltonian $H^0\equiv H^o+ (4\pi/g)Vq$,
with eigenvalue $(4\pi/g)Vq+ (g^3/8\pi q)Vp^2$.

We have just proven that to find a state with charges $p,q$, whose mass
saturates the Bogomol'nyi bound, one must find a normalizable bound state of
the hamiltonian $H^0$, which describes the quantum mechanics on the
(finite-dimensional) $q$-monopole moduli space.

The hamiltonian $H^0$ has been studied in~\cite{blu} and it turns out to
be a N=4 quantum mechanical sigma model on the $q$-monopole moduli
space.

This space is hyperk\"ahler~\cite{at} and has the structure
\beq
M_q=R^3\times {S^1 \times M_q^0 \over Z_q}.
\eeq{6}
Here, $R^3$ denotes the configuration space of the center of mass coordinate
of the monopole, while $S^1$ is labelled by the coordinate $\chi$, with the
identification $\chi \approx \chi + 2\pi$. The metric on $R^3\times S^1$ is
flat, whereas $M^0_q$ is a hyperk\"ahler
{\em non-compact} manifold of real dimension $4(q-1)$~\cite{at}. The total
electric charge of the system is the conjugate variable to $\chi$:
$Q_{el}=-i\partial/\partial\chi$. The discrete group $Z_q$ acts freely on
$S^1$: $z\chi = \chi + 2\pi/q$, and also acts non-trivially on $M^0_q$.
The quantization of the N=4 sigma model on $M_q$ can be done, following
ref.~\cite{sen2}, by quantizing at first on the covering space
\beq
\tilde{M}_q= R^3 \times S^1 \times M^0_q,
\eeq{7}
and then truncating the Hilbert space by projecting on $Z_q$ invariant wave
functions.
Let us write explicitly the hamiltonian $H^0$.
To do this we must explicitly list the moduli,
previously denoted collectively by $m$. They are~\cite{blu}:
\begin{itemize}
\item 4 real
bosonic moduli $Y^a$, $a=1,..,4$ parametrizing $R^3\times S^1$. They
may be denoted equivalently by $\vec{Y}$, $\chi\equiv Y^4$.
\item $2\times 4$ fermionic moduli, $\eta^a_\alpha$, $\alpha=1,2$
related by supersymmetry to the $Y^a$.
\item $4(q-1)$ real
bosonic moduli $X^A$, $A=1,..,4(q-1)$,
parametrizing $M_q^0$.
\item $2\times 4(q-1)$ real fermionic moduli, $\lambda^A_\alpha$,
related by supersymmetry to the $X^A$.
\end{itemize}
The hamiltonian reads
\beq
H^0 = H^1 + H^2, \;\;\;
H^1=-{1\over 2}
g^{q\, ab}{\partial\over \partial Y^a} {\partial\over \partial Y^b} + {4\pi
V\over g}q.
\eeq{8}
The metric on $R^3\times S^1$, as stated before, is flat. Its normalization,
as well as the constant term $(4\pi/g) Vq$,  may be
fixed by demanding that the eigenvalues of $H^1$ reproduce the
nonrelativistic, $g\rightarrow 0$ limit of the mass formula~(\ref{2}):
\beq
g^{q\,ab}=\mbox{diag}\left( {g\over 8\pi Vq},{g\over 8\pi Vq},{g\over 8\pi
Vq},{g^3V\over 8\pi q}\right).
\eeq{8'}
Here we use a real notation to describe the moduli of $R^3\times S^1$ and
$M^0_q$. In this notation, only one of the four supersymmetries of the model is
manifest.
By denoting with $g_{AB}$, $\Gamma_{BC}^A$, and $R_{ABCD}$,
the metric,
Christoffel symbol, and Riemann curvature of $M^0_q$, respectively, and
setting $D_A= \partial_A -
\Gamma_{AC}^Bg_{BD}\bar{\lambda}^C\gamma^0 {\lambda}^{D}$ (notations are as
in~\cite{blu,sen2}),
the hamiltonian $H^2$  reads~\cite{blu,ch}
\beq
H^2= -{1\over 2\sqrt{g}}D_A \sqrt{g}g^{AB}D_{B} -{1\over 12}
R_{ABCD}\bar{\lambda}^A\lambda^C\bar{\lambda}^B\lambda^D.
\eeq{9}
Since the hamiltonian $H^0$ is a sum of two commuting operators, its
eigenfunctions factorize as $\psi=\psi^1 \otimes \psi^2$. $\psi^1$ is an
eigenstate of $H^1$. Notice that $H^1$ is independent of the 8
fermionic coordinates $\eta_\alpha^a$. Thus the eigenvalues of $H^1$ are
16-fold degenerate, and $\psi^1$ is the product of a bosonic wave function,
a plane wave $\exp(iP_aY^a)$, and a fermionic wave function $\xi$. The
fermionic wave function is  a state in the 16-dimensional
fermionic Fock space.
A simple calculation
shows that for each value of $P_a$, the 16 degenerate states correspond to a
4-dimensional, massive spin-one particle,
four spin-1/2 (Majorana), and five real spin-0 particles.
This is exactly the content of a massive short multiplet of N=4
supersymmetry in four dimensions! Moreover,
\beq
H^1\exp(iP_aY^a)\xi=E^1\exp(iP_aY^a)\xi=
\left({g\over 8\pi Vq}\vec{P}^2 + {g^3V\over 8\pi
q}p^2 + {4\pi V\over g}q \right)\exp(iP_aY^a)\xi.
\eeq{9'}
Recalling that $\vec{P}$ is the center of mass momentum of the $q$-monopole,
and that $p\equiv P_4$ is the electric charge, we find that indeed
eq.~(\ref{9'}) gives the non-relativistic limit of the energy of a
Bogomol'nyi state.
The spectrum of $H^2$ is positive definite, by supersymmetry:
$H^2\psi^2=E^2\psi^2$, $E^2\geq 0$.

Since the total energy of a $q$-monopole configuration is $E^1+E^2$, this
implies that Bogomol'nyi states are in one-to-one correspondence with {\em
normalizable} solutions of the equation $H^2\psi^2=0$, in other words, they
are in one-to-one correspondence with the zero-energy eigenvalues of $H^2$.
Notice that for a given value of $q$ and $p\equiv P_4$,
the degeneracy of the eigenvalue $E^1$ already
gives a complete N=4 short multiplet with mass given by eq.~(\ref{2}). Thus,
one would obtain an N=4 short multiplet saturating the Bogomol'nyi bound for
each zero-energy eigenstate of $H^2$. Since at $p=1,q=0$ there is only one
such multiplet, S-duality predicts that a zero energy eigenstate of $H^2$
with charges $p,q$ relatively prime is non-degenerate.

So far, we have mostly recalled the arguments in ref.~\cite{sen2}. Now we
come to the main point of the paper, namely, to prove the existence
of normalizable solutions of the equation
\beq
H^2\psi^2=0.
\eeq{10}

If the bosonic moduli space $M^0_q$ were compact, then the spectrum of $H^2$
would be discrete and one could count straigthforwardly the number of
solutions of eq.~(\ref{10}). One way of doing it would be to perturb the
hamiltonian $H^2$ by adding a N=1
superpotential $W(X)$~\cite{win}.
The superpotential is a real, smooth
function of $X^A$, in terms of which the ordinary potential energy
$V(X)$ reads $V(X)=(1/2)g^{AB}\partial_A W(X)\partial_B W(X)$.
As shown in ref.~\cite{win}, when the spectrum of a supersymmetric
hamiltonian is discrete~\footnotemark,
\footnotetext{Or when, at least, the continuous part of the spectrum is always
larger than a nonzero, positive number, i.e., when the spectrum has a gap.}
the index
$\Delta=N_B-N_F$, i.e. the number of bosonic minus fermionic
zero-energy states of a supersymmetric system, is invariant under a large
class of supersymmetry-preserving deformations of the hamiltonian, which
includes the addition of a superpotential, and can
be computed in the semiclassical approximation.
In the semiclassical approximation, the zero-energy states of the model are
determined by
the points at which the potential $V(X)$ vanishes, i.e. by
the stationary points (denoted here by $X_i$) of the
superpotential (where $\partial W/\partial X^A|_{X_i}=0$ $\forall A$).
If they are isolated and the hessian matrix, $H_{AB\,i}\equiv
\partial^2 W(X)/\partial X^A \partial X^B|_{X_i}$, is non-degenerate,
$N_B-N_F=\sum_i (-)^{\sigma_i}$. Here $\sigma_i$ is the number of negative
eigenvalues of the hessian $H_{AB\,i}$~\footnotemark.
\footnotetext{In one dimension, the sign of $\Delta$ is only a matter of
conventions: one may also set $N_F-N_B=\sum_i (-)^{\sigma_i}$
by redefining the Fock vacuum.}
Moreover, if the hamiltonian is invariant under a symmetry group ($Z_q$, in
our case), the Hilbert space of the model decomposes into irreducible
representations of that group, and one can apply the previous arguments
separately to each representation. Thus, $\Delta^p=N_B^p-N_F^p$, i.e. the
number of bosonic ground states in the representation $p$, minus the
fermionic ones, is an index~\cite{win}.
Unfortunately, $M^0_q$ is noncompact and the spectrum of $H^2$ is gapless.
Nevertheless, we can still
add an F-term perturbation, i.e. a superpotential $(1/n)W$,
where $n$ is a positive integer, and then let $n\rightarrow \infty$.
In this way, for each finite $n$, and by choosing an appropriate $W(X)$,
we can still have a mass gap, and use Witten's argument to prove that there
exists a normalizable zero-energy state for each stationary point of $W$.
Obviously, we must carefully study the fate of those normalizable (bound)
states as $n \rightarrow \infty$. Indeed, nothing guarantees {\em a priori}
that a zero-energy bound state of the perturbed theory, $\psi^2_n$, would
converge to a {\em normalizable} ground state of $H^2$.

To solve this problem, our strategy will be the following:
\begin{enumerate}
\item we will introduce a superpotential $W(X)$,
well defined and nonsingular on the whole $M^0_q$, invariant under $Z_q$,
and with $q$ isolated zeroes such that $\Delta^p\geq 1$, for $p=1,..,q-1$. 
\item We will add the superpotential $(1/n)W(X)$ to the hamiltonian
$H^2$, and let $n\rightarrow \infty$. We will
find that the perturbed ground states $\psi_n^2$ converge pointwise to
\cinf~ solutions, $\psi_\infty$, of the equation
$H^2\psi_\infty=0$.
\item By studying the large-distance behavior of $\psi_\infty$ we will
find that they are normalizable whenever $p$ and $q$ are relatively prime,
{\em as required by S-duality}.
\item Finally, we will exhibit a superpotential obeying the
properties assumed in point 1, and we will comment on 
the uniqueness of the BPS bound states.
\end{enumerate}

Adding a superpotential $(1/n)W(X)$ to $H^2$ modifies the hamiltonian as
follows
\beq
H^2_n=H^2 +{1\over 2n} \nabla_A \nabla_B W \bar{\lambda}^A\lambda^B +
{1\over 2 n^2}
g^{AB}\nabla_A W \nabla_B W\equiv H^2 +\delta H_n^2
\eeq{11}
where $\nabla_A$ is the usual covariant
derivative on $M^0_q$, obtained from the Christoffel connection.

The complete wave function on $M_q$,
$\psi_n=\psi_n^1\otimes \psi^2_n$, is invariant under $Z_q$.
Since the electric charge is $Q_{el}=-i\partial/\partial Y^4$, a state
$\psi_{n\,p}$ of electric charge $p$ is the product of two wave functions
transforming as follows under $Z_q$:
\beq
z\psi^1_{n\,p}=\exp(i2\pi p/q)\psi^1_{n\,p},\;\;\;
z\psi^2_{n\,p}=\exp(-i2\pi p/q)\psi^2_{n\,p}.
\eeq{13''}
Thus, by
denoting with $|0\rangle$ the Fock vacuum of the fermionic Hilbert space,
defined by $\lambda^A|0\rangle=0$ $\forall A$,
one may write the most general solution of the equation $H^2_n\psi^2_{n\,p}=0$
as
\beq
\psi_{n\,p}^2= \sum_{i=0}^{4(q-1)}
\varphi^{{\alpha}_1,{A}_1,...{\alpha}_i,{A}_i}_{n\,p}
(X^A)\prod_{j=1}^i
\bar{\lambda}_{{\alpha}_j}^{_j}|0\rangle, \;\;\;
({A}_l,{\alpha}_l) \neq ({A}_m,{\alpha}_m),\;\;\;
\mbox{for } l\neq m.
\eeq{12}

By substituting formula~(\ref{12}) into the equation $H^2_n\psi^2_{n\,p}=0$,
one can re-write it as a system of $2^{4(q-1)}$ elliptic
second-order partial differential equations. They have the following
structure:
\beq
-{1\over 2}g^{AB}\partial_A{\partial}_{{B}}
\varphi^{{\alpha}_1,{A}_1,...{\alpha}_i,{A}_i}_{n\,p}
+ \mbox{lower derivative terms} =0.
\eeq{13}
This is a special case of elliptic system for which many of the theorems
valid for one-component elliptic equations apply~\cite{mir}.

Now, let us study the fate of $\psi^2_{n\,p}$, as $n\rightarrow\infty$.
First of all, since $\Delta^p\geq 1$, for $p=1,..,q-1$, at finite $n$, there
exists at least one normalizable $\psi^2_{n\,p}$ for each $p=1,...,q-1$.
Since all coefficient of the system~(\ref{13}) are \cinf, the functions
$\varphi^{{\alpha}_1,{A}_1,...{\alpha}_i,{A}_i}_{n\,p}$
are also \cinf.
Obviously, given an arbitrary compact set $K'\subset M^0_q$, one may normalize
the wave function, uniformly in $n$, as follows
\beq
\max_{X\in K', i}\max_{
{\alpha}_1,{A}_1,...{\alpha}_i,{A}_i}
|\varphi^{{\alpha}_1,{A}_1,...{\alpha}_i,{A}_i}_{n\,p}(X)|=1.
\eeq{13'}
Because of this equation, given
a ${\cal C}^0(M^0_q)$ function $f$, and denoting by $\Omega$ the volume of
$K'$, one has
\beq
|\int_{K'}
\varphi^{{\alpha}_1,{A}_1,...{\alpha}_i,{A}_i}_{n\,p}
f|\leq \Omega \max_{K'}|f|.
\eeq{asp1}
In $K'$, the
hamiltonian $H^2_n$ converges pointwise to $H^2$ as $n\rightarrow\infty$
This property, together with a
standard theorem in analysis, ensures that the
$\varphi^{{\alpha}_1,{A}_1,...{\alpha}_i,{A}_i}_{n\,p}|_{K'}$
converge in the weak$^*$ topology to distributions
$\varphi^{{\alpha}_1,{A}_1,...{\alpha}_i,{A}_i}_{\infty\,p}$.
Moreover, since
$H^2_n\psi^2_{n\,p}=0$ for all $n$, one has $H^2\psi_{\infty\,p}^2=0$.
The coefficients of
$H^2$ are \cinf, thus $H^2$ is a hypoelliptic system of differential
equations, and the
$\varphi^{{\alpha}_1,{A}_1,...{\alpha}_i,{A}_i}_{\infty\,p}
(X)|_{K'}$ belong indeed to ${\cal C}^\infty(K')$~(see
for instance chapter IV of ref.~\cite{ell}).

Since the previous convergence result are valid on an
arbitrary compact set $K'$,
we find that the coefficients
$\varphi^{{\alpha}_1,{A}_1,...{\alpha}_i,{A}_i}_{n\,p}(X)$
of a charge-$p$ wave function $\psi^2_p$
are the point-wise limit
\beq
\varphi^{{\alpha}_1,{A}_1,...{\alpha}_i,{A}_i}_p(X)=
\lim_{n\rightarrow\infty}\sum_{k=0}^{(q-1)}\exp(i2\pi pk/q)
\varphi^{{\alpha}_1,{A}_1,...{\alpha}_i,{A}_i}_n(z^kX).
\eeq{13'''}
Thus,
eq.~(\ref{13'''}) proves that, for each value of $p=1,...,q-1$,
there exists at least one zero-energy
wave function $\psi^2_p$. The coefficients
$\varphi^{{\alpha}_1,{A}_1,...{\alpha}_i,{A}_i}_p(X)$ of
this wave function are \cinf, but they are not necessarily bounded
on $M^0_q$.
In other words, not
all these $\psi^2_p$ correspond to bound states: only the normalizable
ones do; our next task is to find them.

Since the coefficients of $\psi^2_p$ are smooth, we need only study the
asymptotic behavior of them. To do this we must understand the structure of
the asymptotic regions of $M^0_q$.
This can be done as follows: the space $M_q$ is a
desingularization of the symmetric product of $q$ one-monopole spaces
$R^3\times S^1$~\cite{at}. Its asymptotic regions correspond to
configurations where these $q$ monopoles can be split into two well
separated sets: a set of
$k$ monopoles, with center of mass $\vec{Y}_k$,
and a set of $q-k$ monopoles, with center of mass $\vec{Y}_{q-k}$.
In this case, since a multi-monopole configuration is localized in
space~\footnotemark,
\footnotetext{This is most easily seen in the singular gauge.}
one can write the charge-$q$ solution $\phi^o_q(\vec{x},m_q(t))$ (cfr.
eq.~(\ref{5'}))  as an approximate linear superposition
\beq
\phi^o_q(\vec{x},m_q(t)) \approx \phi^o_k(\vec{x},m_k(t)) +
\phi^o_{q-k}(\vec{x},m_{q-k}(t))   .
\eeq{14}
Moreover, since the action $S[\phi(t,\vec{x})]$ itself is local, one finds
\beq
S[\phi^o_{q}(\vec{x},m_q(t))] \approx S[\phi^o_{k}(\vec{x},m_k(t))] +
S[\phi^o_{q-k}(\vec{x},m_{q-k}(t))].
\eeq{14'}
This equation implies that the metric of the moduli space factorizes when
$d\equiv|\vec{Y}_k-\vec{Y}_{q-k}| \rightarrow \infty$:
\beq
\lim_{d\rightarrow\infty} g^q_{IJ}=
\mbox{diag}(g^k_{I_1J_1},g^{q-k}_{I_2J_2}),\;\;\;
I_1,J_1=1,...,4k,\;\;I_2,J_2=4k+1,...,4q.
\eeq{14''}
Equivalently, asymptotically $\tilde{M}_q$ factorizes (locally)
into $\tilde{M}_k\times \tilde{M}_{q-k}$. More precisely, 
by denoting with $U^q_d$ a neighborhood of a
point $m_q\in \tilde{M}_q$, such that $|\vec{Y}_k-\vec{Y}_{q-k}|=d$, the
following local factorization property holds:
$\lim_{d\rightarrow\infty}U^q_d=U^k\times U^{q-k}$. Here $U^k$ and $U^{q-k}$
are neighborhoods of $\tilde{M}_k$ and $\tilde{M}_{q-k}$, respectively.
It should be possible to prove rigorously these properties using the results of
ref.~\cite{tau}.

Eq.~(\ref{14''}) implies that the hamiltonian $H^1 + H^2_n$ also factorizes
asymptotically. To study it let us parametrize $\tilde{M}_k=R^3\times
S^1\times M^0_k$ with the
following coordinates: $Y^a_k$, parametrizing $R^3\times S^1$, and $X^{A_1}_k$,
parametrizing $M^0_k$. Let us similarly parametrize $\tilde{M}_{q-k}$ by
$Y^a_{q-k}$, $X^{A_2}_{q-k}$, and define
\bea
\vec{Y}&=&{k\over q} \vec{Y}_k + {q-k\over q} \vec{Y}_{q-k} \nonumber \\
\vec{Z}&=& \vec{Y}_k - \vec{Y}_{q-k}.
\eea{15}
The factorization of the metric given in eq.~(\ref{14''}) then implies that,
when $d\equiv |\vec{Z}| \rightarrow \infty$  the asymptotic form of the
equation for $\psi_{n\,p}$ is
\bea
&&\left({4\pi\over g}Vq + {g\over 8\pi V q}\vec{P}^2 + {g^3V\over 8\pi
q}p^2\right)\psi_{n\,p}(\vec{Z})= (H^1+H^2_n)\psi_{n\,p}(\vec{Z}) \approx
\nonumber \\
&&\approx \left(-{1\over 2}
g^{k\, ab}{\partial\over \partial Y^a_k} {\partial\over \partial Y^b_k}
 -{1\over 2} g^{q-k\, ab}{\partial\over \partial Y^a_{q-k}}
{\partial\over \partial Y^b_{q-k}} + {4\pi V\over g}q + 
+ H^2_k + H^2_{q-k} + \delta H^2_n
\right)\psi_{n\,p}(\vec{Z}).\nonumber \\ &&
\eea{16}
Here we have emphasized the dependence of $\psi_{n\,p}$
on $\vec{Z}$ (i.e. the
relative distance between the two multi-monopole clusters).
The metrics $g^{k\, ab}$ and $g^{q-k\, ab}$ are as in eq.~(\ref{8'}),
whereas the hamiltonians $H^2_k$, $H^2_{q-k}$ are as in eq.~(\ref{9}).
Equation~(\ref{16}) also implies that asymptotically
the wave function $\psi_{n\,p}(\vec{Z})$ is
a linear superposition of vectors $\exp(i
P_aY^a)\Psi_{n\,p_1,p_2,\mu}(\vec{Z})$, where
\bea
(H^2_k + H^2_{q-k}+ \delta H^2_n)\Psi_{n\,p_1,p_2,\mu}(\vec{Z})
&=&\mu \Psi_{n\,p_1,p_2,\mu}(\vec{Z}), \nonumber \\
-i{\partial \over \partial Y^4_k} \Psi_{n\,p_1,p_2,\mu}(\vec{Z}) &=& p_1
 \Psi_{n\,p_1,p_2,\mu}(\vec{Z}),  \nonumber \\
-i{\partial \over \partial Y^4_{q-k}} \Psi_{n\,p_1,p_2,\mu}(\vec{Z}) &=& p_2
 \Psi_{p_1,p_2,\mu}(\vec{Z}).
\eea{17}
Here $p_1,p_2 \in Z$,
$p_1+p_2=p$ and $\mu \geq 0$. This last property holds because 
the spectrum of the supersymmetric hamiltonian $H^2_k + H^2_{q-k}$ is 
non-negative and, away from the stationary points of the potential, where the
semiclassical approximation holds, 
$\delta H^2_n\approx (1/2n^2)g^{AB}\nabla_A W \nabla_BW\geq 0$ .
Substituting eq.~(\ref{17}) into eq.~(\ref{16}) one finds
\beq
\left( -{gq\over 8\pi V k(q-k)}{\partial \over \partial \vec{Z}}
{\partial \over \partial \vec{Z}} + {g^3V\over 8\pi k}p_1^2 + {g^3V\over 8\pi
(q-k)} p_2^2 - {g^3V\over 8\pi q} p^2 +\mu
\right)\Psi_{n\,p_1,p_2,\mu}(\vec{Z})=0.
\eeq{19}
This is an equation in $\vec{Z}$ whose normalizable solutions behave
asymptotically as a Yukawa potential:
\beq
\Psi_{n\,p_1,p_2,\mu}(\vec{Z})\approx \mbox{const}\,
|\vec{Z}|^{-1}\exp(-M_n(p_1,p_2,k,\mu)|\vec{Z}|),
\eeq{21}
where
\beq
M_n(p_1,p_2,k,\mu)=V \sqrt{{g^2(q-k)\over q}p_1^2 + {g^2 k\over q} p_2^2 -
{g^2 k(q-k)\over q^2} p^2 + {8\pi k(q-k)\over gqV}\mu}.
\eeq{22}
This equation gives a bound, {\em independent of $n$}, on the asymptotic
behavior of the wave function $\psi_{n\,p}$:
\beq
\psi_{n\,p}(\vec{Z}) \leq
\mbox{const}\, |\vec{Z}|^{-1} \exp [-M(k)|\vec{Z}|],
\eeq{23}
where
\beq
M(k)= \min_{p_1,p_2\in Z}M_{\infty}(p_1,p_2,k,0).
\eeq{24}

Thus, when $M$ is strictly positive, eq.~(\ref{23}) guarantees that the wave
function $\psi_{n\,p}(\vec{Z})$ is normalizable in $\vec{Z}$, uniformly in
$n$. If $M(k)>0$
for any $k=1,..,q-1,$ then the wave function
is {\em uniformly} bound by a normalizable wave function
in all asymptotic regions of $M^0_q$. Since the bound is uniform in $n$, the
wave function $\psi_p$ is also normalizable.

 {\em
Now, we have reduced the problem of finiding zero-energy bound states of
$H^2$ with charge $p$ to the problem of finding a pair of charges $(p,q)$
such that $\min_{k=1,..,q-1}M(k)>0$.}

An inspection of eq.~(\ref{22}) shows that $M^2(p_1,p_2,k,0)$ is a positive
semidefinite quadratic form in $p_1,p_2$ which vanishes at $p_1/k=p_2/(q-k)$.
As we already pointed out, this is impossible if $p$
and $q$ are relatively prime. Thus, whenever $p$ and $q$ are relatively
prime, there exists at least one zero-energy bound state of $H^2$: this
is consistent with the prediction of S-duality!

At this point, we must exhibit a superpotential $W(X)$ with the desired
properties, and comment on the arbitrariness involved in the
procedure so far described. The choice of $W(X)$ is far from
unique: as we have just seen, the bound on the asymptotic behavior of the
wave function is independent on the asymptotic 
form of the perturbing superpotential. Thus, to complete our proof 
we only need to exhibit explicitly a superpotential obeying the conditions
in point 1 above. To do this, we use the representation of $M_q^0$ given in 
ref~\cite{at}. There, $M_q^0$ was represented by the set of 
coefficients $(a,b)\in C^{2q-1}$ 
obeying the algebraic constraint $\Delta(a,b)=1$, where
$\Delta(a,b)$ is the resultant
\beq
\Delta(a,b)= \det\left( \begin{array}{lllllllll} a_0 & a_1 & ... & ... &
a_{q-1} & & & & \\ & a_0 & ... & ... & ... & a_{q-1} & & & \\
& & ...& ... & ... & ... & & & \\ & & & & a_0 & ... & ... & ... & a_{q-1}\\
b_0 & b_1 & ... & b_{q-2} & 0
& 1 & & & \\ & b_0 & ... & ... & b_{q-2} & 0 & 1 & & \\
& & ... & ... & ... & ... & ... & & \\
& & & b_0 & ... & ... & b_{q-2} & 0 & 1 \end{array} \right).
\eeq{1r}
We set
\beq
W(a,b)=\kappa \sum_{i=1}^{q-1}\left({|a_i|^2\over \sum_{l=0}^{k-1} |a_l|^2}
+
|b_{i-1}|^2\right).
\eeq{2r}
Here $\kappa$ is an arbitrary nonzero constant~\footnotemark.
\footnotetext{Notice that by choosing the 
constant $\kappa$ large enough, we can insure
that the semiclassical approximation holds to arbitrary accuracy
everywhere outside of the stationary points.}
This function is symmetric under $Z_q$ and it has 
exactly $q$ isolated stationary points at
\beq
a_i=b_{i-1}=0,\;\;\; i=1,..,q-1,\;\;\; a_0=\exp(2\pi i r/q),\;\;\; r=0,..,q-1.
\eeq{0rr} 
To prove this, we introduce a Lagrange multiplier $\lambda$
which implements the constraint $\Delta(a,b)-1=0$ and we write the
stationarity conditions subject to the constraint.
\bea
{\partial  \over \partial \bar{a}_k}\{\lambda[\Delta(a,b)-1]+
\bar{\lambda}[\bar{\Delta}(a,b)-1]+ 
W(a,b)\}&=&0,\nonumber \\
{\partial  \over \partial \bar{b}_k}\{\lambda[\Delta(a,b)-1] +
\bar{\lambda}[\bar{\Delta}(a,b)-1] +
W(a,b)\}&=&0,\nonumber \\ 
{\partial  \over \partial a_k}\{\lambda[\Delta(a,b)-1] + 
\bar{\lambda}[\bar{\Delta}(a,b)-1]+
W(a,b)\}&=&0,\nonumber\\
{\partial  \over \partial b_k}\{\lambda[\Delta(a,b)-1] + 
\bar{\lambda}[\bar{\Delta}(a,b)-1]+
W(a,b)\}&=&0,\nonumber \\
\Delta(a,b)&=&1.
\eea{1rr}
Since $W(a,b)$ and $\Delta(a,b)$ are homogeneous functions in $a_k$, of
degree $0$ and $q$ respectively, 
the first equation in~(\ref{1rr}), multiplied by $a_k$ and summed over $k$,
gives $\lambda=0$. The second equation then
implies $b_i=0$, $i=0,...,q-2$. At $b_i=0$, $\Delta(a,b)=a_0^q$, and the
constraint implies $a_0^q=1$ i.e. $|a_0|=1$; thus, the third equation  gives  
$a_i=0$, $i=1,...,q-1$.

There are many other ways to construct superpotentials
with the required properties\footnotemark. These superpotentials
break the $N=4$ supersymmetry of the sigma model to $N=1$. In $N=1$
supersymmetry, we know how to control the difference $N^p_B-N^p_F$, but not
$N^p_B$ or $N^p_F$ separately; thus, there exists the (unlikely)
possibility of finding extra zero-energy
normalizable states with $p$, $q$ relatively
prime. 
\footnotetext{The superpotentials $W$ are not completely arbitrary, though. 
Indeed, since the 
action of $Z_q$ on $M_q^0$ is free~\cite{at}, their stationary points always
fall into orbits of order $q$; thus, $\Delta^p\geq 1$ for $p=0,...,q-1$.}
\vskip .1in
\noindent
Acknowledgements
\vskip .1in
\noindent
L. Caffarelli, S. Cappell, W. Fulton, G. Gibbons,
J. Harvey, G. Moore, and A. Sen are acknowledged for 
advice and references. The Aspen Center for Physics is also 
acknowledged for providing a stimulating research environment during the
workshop on {\em Non-Perturbative Supersymmetry and Strings}, July 3-29, 1995.
This work was supported in part by NSF Grant PHY-9318781.


\begin{thebibliography}{99999}
\bibitem{misc} A. Sen, Int. Jou. Mod. Phys. A9 (1994) 3707;
J.H. Schwarz and A. Sen, Nucl. Phys. B411 (1994) 35; Phys.
Lett. B312 (1993) 105; C. Vafa and E. Witten, Nucl. Phys. B431 (1994) 3;
A. Giveon, L. Girardello, M. Porrati and A. Zaffaroni, Phys. Lett. B334
(1994) 331; Nucl. Phys. B448 (1995) 127.
\bibitem{sen1} A. Sen, Int. Jou. Mod. Phys. A8 (1993) 2023.
\bibitem{sen2} A. Sen, Phys. Lett. B329 (1994) 217.
\bibitem{mo} C. Montonen and D. Olive, Phys. Lett. B72 (1977) 117.
\bibitem{osb} H. Osborn, Phys. Lett. B83 (1979) 321.
\bibitem{gno} F. Englert and P. Windey, Phys. Rev. D14 (1976) 2728; P.
Goddard, J. Nuyts and D. Olive, Nucl. Phys. B125 (1977) 229.
\bibitem{bog} E.B. Bogomol'nyi, Sov. Jou. Nucl. Phys. 24 (1976) 449.
\bibitem{wo} E. Witten and D. Olive, Phys. Lett. B78 (1978) 97.
\bibitem{se} G. Segal, unpublished.
\bibitem{w} E. Witten, Phys. Lett. B86 (1979) 283.
\bibitem{gm} G. Gibbons and N.S. Manton, Nucl. Phys. B274 (1986) 183.
\bibitem{blu} J.P. Gauntlett, Nucl. Phys. B411 (1994) 443; J. Blum, Phys.
Lett. B333 (1994) 92.
\bibitem{at} M. Atiyah and N. Hitchin, {\em The Geometry and Dynamics of
Magnetic Monopoles}, Princeton Univ. Press, Princeton NJ (1988)
\bibitem{ch} M. Claudson and B. Halpern, Nucl. Phys. B250 (1985) 689.
\bibitem{win} E. Witten, Nucl. Phys. B202 (1982) 253.
\bibitem{mir} C. Miranda, {\em Partial Differential Equations of Elliptic
Type}, Springer-Verlag, Berlin (1970).
\bibitem{ell} N. Shimakura, {\em Partial Differential Operators of Elliptic
Type}, Translations of Mathematical Monographs, AMS, Providence RI (1992).
\bibitem{tau} C.H. Taubes, Comm. Math. Phys. 97 (1985) 473.
\bibitem{wb} J. Wess and J. Bagger, {\em Supersymmetry and Supergravity},
Princeton Univ. Press, Princeton NJ (1992).
\end{thebibliography}
\end{document}